\documentclass{icrc}

\usepackage{times}
 \usepackage{graphicx} 

\begin{document}

\title{Very High Energy $\gamma-$ ray Emission from Crab and Geminga Pulsars}
\author[1]{P.R.Vishwanath}
\author[1]{B.S.Acharya}
\author[1]{P.N.Bhat}
\author[1]{V.R.Chitnis}
\author[1]{P.Majumdar}
\author[1]{M.A.Rahman and B.B.Singh}
\affil[1]{Tata Institute of Fundamental Research,Colaba\\
Mumbai,400005, India}

\correspondence{vishwa@tifr.res.in}

\runninghead{Vishwanath et al: VHE $\gamma-$ rays from Crab and Geminga Pulsars}
\firstpage{1}
\pubyear{2001}


\maketitle

\begin{abstract}
The main inference from the experiments of the '80s that the time-averaged energy 
spectra  of pulsars had to steepen in the GeV-TeV energy region has been reinforced in 
the '90s from experiments with higher  sensitivities. However, results from several 
experiments from the past and the more sensitive experiments at present can be 
reconciled by invoking a  possibly different component  arising in the TeV region. The 
results of the preliminary analysis of the data being collected with the PACT array 
will be 
presented. 
\end{abstract}

\section{Introduction}
  Almost all the VHE $\gamma-$ ray groups looking for VHE $\gamma-$ ray emission 
from  pulsars in the '70s did find some signals from the Crab pulsar, but the results could 
not be repeated.  While there were several possible reasons for non-reproducibility of the 
results , it was becoming clear from  later experiments in '80s that time averaged pulsar 
spectra have to steepen at the energies 500 to 1000 GeV. As we will see below, these 
tentative conclusions from that era have been borne out in general by the more recent 
and sensitive VHE $\gamma-$ ray experiments, with both imaging and non-imaging 
techniques.
Pulsar studies have become  more interesting for the VHE gamma ray astronomer 
since the EGRET results which detected seven $\gamma-$ ray pulsars  
[\cite{Th20}].  At $>$ 1 GeV energies, the light curves change with the dominance of 
either the main or the inter pulse. Further the spectra show dominant power at energies of 
$\leq$ 1 GeV with  a falloff at higher energies. All these features  indicate  changes in 
characteristics at higher energies. Thus, the VHE $\gamma-$ ray experiments have the 
interesting task of finding the cut-off energies for the emission mechanism acting at the 
MeV-GeV energy region in the environs of the pulsars.
The Outer Gap(OG) models for pulsar emissions predict an inverse Compton 
component due to gap accelerated particles  with a peak around 1 TeV. While there are 
variations in the OG models, all models predict a TeV emission component. Thus,  
while the emission process giving rise to MeV-GeV gamma rays could peter out at sub-
TeV energies, it is important to detect other processes in pulsars which could give rise 
to GeV-TeV $\gamma-$ rays.  While the recent trend in the field has been to lower the 
threshold energies, it is interesting that one has really to look at the higher energy part of 
the data to detect this emission. Preliminary results from a new experiment  which has 
started out at Pachmarhi will be specially looked at for evidence for these higher energy 
emissions.  Recent reviews on VHE gamma ray emission from pulsars can be found in 
Fegan [\cite{Fe96}] and Kifune [\cite{Kif96}].

\section{Earlier Results on Crab and Geminga Pulsars}

   While there have been many observations in the 1970s with and without 
absolute phase information, we will restrict ourselves to the results in the 80s which in 
general had absolute phase information. The Durham group, working at the Dugway site 
in USA, detected a 4.3$\sigma$  signal at the main pulse phase position from 103 hours of 
observation at an  energy threshold of 1 TeV.  The flux was found to be   
$5 \times 10^{-12}$ $sec^{-1}$ $cm^{-2}$. It should be noted that the detected flux is quite small, 
about 3 per 
hour. With arrival direction estimation, they also showed that the signal falls off away 
from the pulsar direction. The Tata group, earlier at Ooty and later at Pachmarhi, had 
many hours of observation on the pulsar and detected transients but no time-averaged  
signal. As a representative of the Tata observations, one can use the  Ooty results of 3 
years (a total of 105 hours of observation), all with absolute phase, which did not show a 
signal  at the main peak and where an upper limit of 4.6 $\times 10^{-12}sec^{-1}cm^{-2}$ 
for Energy $>$ 800 GeV was derived for the time averaged emission.[\cite{vishwa87}] 
The Asgat experiment, with 50 hours of exposure, derived an upper limit of 
$2.2 \times 10^{-12}$ $sec^{-1}$ $cm^{-2}$ at $>$600 GeV[\cite{Gor93}]. 
Recently, a thorough search for 
pulsed emission has been done with the Whipple imaging telescope.[\cite{Bur99}] 
Using the data collected between 1995 and 1997, for an exposure of about 73 hours, 
they give an upper limit for pulsed emission to be $4.8 \times 10^{-12}sec^{-1}cm^{-2}$
and $1.2 \times 10^{-12}sec^{-1}cm^{-2}$ 
above $>$ 250 and 1000 GeV respectivley. 
HEGRA experiment has also given only upper limits for emission of VHE $\gamma-$ rays from 
Crab abd Geminga pulsars. [\cite{Aha99}]
The CELESTE experiment and 
the STACEE experiments, both with solar arrays, also do not see evidence for pulsed 
emission in their respective energy regions. 
\par After the mystery that Geminga had created in the '80s was resolved by the 
EGRET experiment in the '90s, two groups [\cite{Bow93}],[\cite{vishwa93}]  
found modest pulsar signatures in their archival data  with VHE $\gamma-$ ray peaks 
coinciding with the peaks in the low energy  data. The flux reported by the Ooty group( 
with simultaneous observation at 2 sites for a total of 28 hours ) at  energy thresholds of 
0.8 and 1.7 TeV was $2.1 \times 10^{-11}sec^{-1}cm^{-2}$ and $0.44 \times 10^{-11}sec^{-1}cm^{-2}$ 
respectively. Later, in the 
'90s, an upper limit for pulsed emission from Geminga was given  by the Whipple 
imaging telescope.[\cite{Aker93}] 
\par During 1992-1994, the Tata group working at Pachmarhi ran an interim array for 
testing some of the hypothesis concerning Lateral Distribution(LD) (discussed below ) of  
VHE $\gamma-$ rays.  8 banks of mirrors, each of total area $2.5m^{2}$  were 
deployed in a 80m $\times$ 100m area. The exposures for Crab and Geminga were 80 and 49 
hours respectively [\cite{vishwa97}]. It was shown that the events at phases at which 
GeV  emission was found(Main pulse region for Crab and around 0.6 for Geminga) 
displayed the same features as expected from gamma ray events. The mean Cerenkov 
light (measured in ADC units) per event was also higher at these phases, an indication of 
the gamma ray admixture. The number of events at the Main Phase region for the Crab 
was about 3.5 per hour.




\section{The new results from PACT}

A totally new set up  PACT has been commissioned recently at Pachmarhi with 
25 telescopes, each with 7 mirrors , having a total area of $105m^{2}$ of mirrors 
[\cite{Bha20}] for using both the LD aspects and the arrival direction information for the 
increase of sensitivity.  A total of 45 hours of data and 20 hours of data on Crab and 
Geminga pulsars were taken respectively. While the Crab data ranged from runs in  Nov 
1999 to Nov 2000, all the Geminga data was taken in Dec 2000. While the ephemeris for 
Crab was taken from the data provided by the Jodrell Bank, the latest parametrization of 
the EGRET group for the pulsar period was used for Geminga. Standard algorithms were 
used to get the phase for each event. Fig 1(a) shows the Crab pulsar phasograms without any 
cut and with the cuts. It can be seen that the phasogram without any cut does not show 
any significant emission at any phase. Therefore, an upper limit to the time averaged 
emission is derived above $>$900 GeV as $2 \times 10^{-12}sec^{-1}cm^{-2}$ This upper limit 
is similar to those set by most experiments (Results on Geminga will be presented later in 
the conference).
\par Many simulations, including the recent extensive calculations by Chitnis and 
Bhat [\cite{Chi99}] have shown that the gamma ray lateral distribution (LD) is 
very different from that of protons in having an almost constant density till about 120 
meters where it becomes higher resulting in the so called hump region (a result of the 
focusing property of Cerenkov radiation) for the next 20 meters. The fall off from the 
hump region is similar to the monotonic decrease for the proton showers from the core. 
The Monte Carlo simulations with these inputs for the PACT array showed that a gamma 
ray event would have at large values of total Cerenkov pulse height since a typical 
gamma ray would deposit much more Cerenkov light than a typical cosmic ray. Further, 
it was seen that $\beta$, a LD parameter which is a measure of how large the detector 
with the maximum 
pulse height would be compared to the rest of the detectors, would be useful for 
distinguishing between proton and gamma ray showers. The details are discussed in the 
conference papers of the PACT experiment.[\cite{vishwa01}]
\par The experiment is capable of good angular resolution. Cuts were applied 
on the space angle difference between the event and the pulsar direction. Further, a cut 
was imposed on the number of mirrors hit and the $\beta$ parameter both of which place the  
events at higher energies. The peak seen at the Main Pulse phase in the lower half of Fig. 
1 is a 4 $\sigma$ signal. Fig 2 shows the rate per hour of the events in the Main Pulse region 
for the 3 sub periods (1999 Nov, 2000 Jan, 2000 October and 2000 November) of the 
data taking. Within errors, the rate is same for the 3 data periods. The mean rate  is 3.4 
$\pm$ 1.2 per hour which compares well with the detected rate with the interim array with 
similar cuts  discussed above.

\begin{figure}[t]
\includegraphics[width=8.3cm]{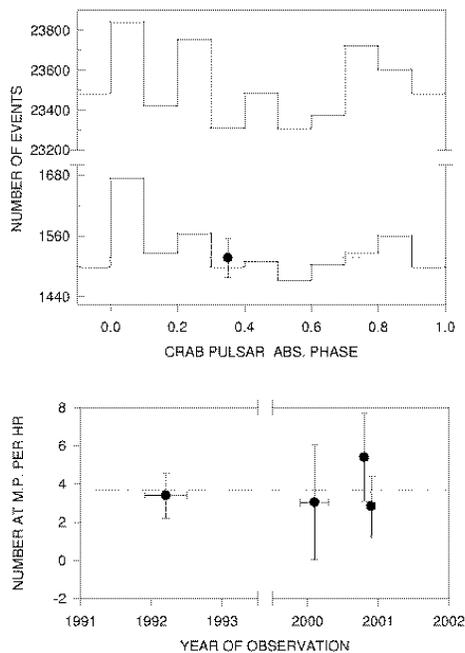} 
\caption{(a)The phasogram of events from Crab pulsar during 1999-2000. 
The top part is the
phasogram of all events without any cuts. The bottom part is the result of several cuts
(see text).
(b) The rate of detected events in the Main Pulse region. The data are from the 1992
Interim array and  the recent PACT experiment}
 
\end{figure}

\section{Discussions}

It is certain that most of the pulsar energy spectra have to  steepen in the energy 
range of the atmospheric Cerenkov experiments. The precise energy region of the 
steepening will be clear when some of the lower energy threshold experiments become 
fully functional. By locating the cut-off energy, the experiments can help distinguish 
between the contending models for gamma ray emission by pulsars.  Polar Cap models 
predict steep spectral cut-offs due to magnetic pair production attenuation and essentially 
no detectable emission above about 50 GeV. The spectral cut-off predicted in OG models 
is more gradual.
\par Some comments have to be made about the results from the earlier generation
experiments. Lacking detailed Monte Carlo calculations which are regular features of the
present day experiments, the energy thresholds and collection area estimates would have
been less precise. However, the calculations in getting the pulsar phase for the event had
been well known. Further, when at least two experiments see the same effect at the same
phase, the credibility of the overall effect should be  quite high.
Some remarks are also in order
about the non-reproducibility which plagued experiments in the early days when the
energy thresholds of  experiments were varying with constant attempts to lower the
threshold. If a new component does surface at $>$ 1 TeV energies, it is not surprising that
there were conflicting results depending on the energy threshold of the experiment.
\par It is seen that the recent results from the Pachmarhi experiment do indicate a finite
number of events  from the Crab Pulsar. While the cuts need to be understood properly
before an actual flux could be given, it would be obviously greater than the detected flux
which is at the level of $7 \times 10^{-13}sec^{-1}cm^{-2}$. It should also be noted that the
cuts for the data place the events detected at energies $>$ 1.5 to 2 TeV, much greater than
the actual  energy threshold of the experiment. The fact that similar flux was seen from
the interim array is also interesting.
Thus, it is possible that
the earlier results from Durham as well as the new preliminary results from Pachmarhi do
point to a new pulsar component at higher energies.
\par Long exposures are very necessary to detect such flux levels. Detected flux from
the Crab pulsar is 2-3 per hour in the experiments of Dowthwaite et al and the recent
results from the Pachmarhi experiment. This has to be compared to the flux levels of 2-3
per minute from the whole nebula. If there is a finite flux from the Crab pulsar at TeV
energies, it should show up as a mild bump in the overall energy spectrum from Crab. If
the nebular emission at few hundred GeVs is solely due to inverse Compton scattering
process, a new process at TeV energies should add to the nebular flux. One should
however keep in mind the very low flux from pulsar and only experiments with good
energy resolution will be able to opine on this issue.





%
\begin{figure}[t]
\end{figure}

%




\begin{acknowledgements}
It is a pleasure to thank Sarvashri A.I.DSouza, J.Francis, 
K.S.Gothe, 
B.K.Nagesh,
M.S.Pose, P.N.Purohit, K.K.Rao, S.K.Rao, S.K.Sharma, A.J.Stanislaus,
P.V.Sudershanan, S.S.Upadhya, B.L.Venkatesh Murthy for their participation 
in various
aspects of the experiment.

\end{acknowledgements}




\end{document}